\documentstyle[12pt]{article}
\begin{document}
\setlength{\unitlength}{1mm}\thicklines
\tolerance=2000
\hbadness=2000
\def \O {\Omega}
\def \e {\epsilon}
\def \s {\sigma}
\def \i {\infty}

\vspace{2cm}
\Large
\begin{center}
TRANSFER MATRIX METHOD\\
AND INTERMITTENCY GENERATING DYNAMICS
\end{center}

\vspace{2cm}
\normalsize
\begin{center}
$A.V.Batunin^1$, $S.M.Sergeev^2$\\
\end{center}

$^1$ - Institute for High Energy Physics,Protvino,142284,Russia;\\
{}~~$^2$ - Branch of Novosibirsk Institute of Nuclear Physics,Protvino,\\
{}~~~142284,Russia.

\vspace{1cm}
\begin{center}
Abstract
\end{center}
A transfer matrix method relating the process
of refinement of a fractal measure to thermodynamic formalism
of an appropriate Ising model is applied to the analysis of
intermittency in hadron collisions revealing that underlying dynamics
is that of period doublings.
\end{center}

\newpage
\section{Introduction}

Fractals appear in hadron physics as self-similar distributions
of particles generated in high energy collisions [1]. Quantitatively,
these self-similar structures are usually characterized by
the exponents $\phi_q$ of the power-like rise of factorial moments
as a bin size of the corresponding phase space decreases [2].
The most common are partitions along the rapidity axis or
two-dimensional ones in rapidity and azimuthal angle.
The quantities $\phi_q$ are related with the generalized
Renyi dimensions $D_q$, whose infinite set describes a given
fractal measure [3]. The description by $D_q$-s (or by the spectral function
$f(\alpha)$ linked to $D_q$-s via Legendre transforms)
is customary anticipated to
be ``static'', i.e., says nothing about the dynamics that gives
rise to this measure.

However, Feigenbaum, Jensen, and Procaccia (FJP) [4] have shown
that the dynamical process that is responsible for the construction
of the measure can be read from its static description.
The key idea of FJP method rests on thermodynamic formalism of
dynamical systems that maps the process of refinement of the
fractal measure to a transfer matrix theory of an appropriate
Ising model. FJP method was already successfully applied to the
analysis of the appearance and development of turbulence in fluid,
in particular, in a forced Reyleigh-Benard
experiment using mercury as the fluid [4].

Below we briefly describe FJP method and apply it to the analysis
of events in hadron-hadron collisions. Such analysis shows that
the dynamics underlying experimental rapidity distributions
of charged particles is that of period-doubling bifurcations
of the corresponding phase space trajectories [5] corroborating
the hypothesis on the Feigenbaum universality in hadron
multiproduction proposed earlier [6,7] by one of us (A.V.B.).

\section{FJP method}
Let $\cal S$ be a set with a measure $d\mu$ defined on the
continuum domain $\bar S$. For a given $N$, let a partition
$\{\Omega_i\}$, $\Omega\in\bar S$, $\Omega_i\cap\Omega_j = \emptyset$,
be a cover of the set $\cal S$ so that
${\cal S}\subset\cup_{i=1}^{N}\Omega_i$. For any given
$\Omega_i$ let $l_i$ be its diameter and
$p_i = \int_{\Omega_i}d\mu$ be the measure of $\Omega_i$,
$\sum_{i=1}^{N} p_i = 1$. Define the corresponding partition
function $\Gamma (q, \tau)$,
\begin{equation}
\Gamma(q,\tau) = \sum_{i=1}^{N} \frac{p_i^q} {l_i^{\tau}}
\end{equation}
For large $N$, $\Gamma (q, \tau)$ is of the order of unity only when
\begin{equation}
\tau(q) = (q-1)D_q,
\end{equation}
where $D_q$ are the generalized Renyi dimensions. If $\tau > \tau (q)$
then $\Gamma (q, \tau) = \infty$, if $\tau < \tau (q)$
then $\Gamma (\tau, q) = 0$ as $l \rightarrow 0$. A convenient way
of calculating $\tau(q)$ is therefore to fix
$\Gamma (q, \tau)$ to a number (say, $\Gamma (q, \tau )$ = 1) as
the partition is refined. We consider special partitions such that
$p_i = p = 1/N$. Then we get
\begin{equation}
N^{q(\tau)} = \sum_{i=1}^{N}l_i^{-\tau},
\end{equation}
and
\begin{equation}
q(\tau) = \frac{log \sum_{i=1}^{N}l_i^{-\tau}} { log N} .
\end{equation}
Consider now a sequence of $N=N_n=a^n$; $a,n$ positive integer.
For any $n$ we can label $l_i$ ($1 \leq i \leq N$) as $l_{\e_n,...,\e_1}$,
where $\e_i $ takes on values $0,1, ...,a-1$. Next define the
daughter-to-mother ratio [8]
\begin{equation}
{l_{\e_{n+1},\e_n,...,\e_1}\over l_{\e_n,...,\e_1}}
= \s_{\e_{n+1},\e_n,...,\e_1}.
\end{equation}
Suppose this ratio depends only on the two latter indices $\e_{n+1},\e_n$:
\begin{equation}
\s_{\e_{n+1},\e_n,...,\e_1} = \s_{\e_{n+1},\e_n}.
\end{equation}
Obviously, $\s_{\e_{n+1},\e_n}$ defines a (transfer) matrix $a \times a$.
Then, any diameter can be written as the product of the transfer matrix
elements
\begin{equation}
l_{\e_n,...,\e_1} = \s_{\e_n\e_{n-1}}\s_{\e_{n-1}\e_{n-2}}...
\s_{\e_2 \e_1}l_{\e_1}
\end{equation}
(with no summation), and eq.(3) becomes
\begin{equation}
a^{nq(\tau)} = \sum_{\e_n,\e_1} [\s(\tau)]^n_{\e_n\e_1}l_{\e_1}^{-\tau},
\end{equation}
where
\begin{equation}
[\s(\tau)]_{\e_2\e_1} = \s_{\e_2\e_1}^{-\tau}.
\end{equation}
In the thermodynamic limit ($n\rightarrow\infty$) we obtain
\begin{equation}
a^{q(\tau)} = \lambda_+(\tau),
\end{equation}
where $\lambda_+(\tau)$ is the largest eigenvalue of the matrix $\s(\tau)$.
Obviously,
\begin{equation}
q_a(\tau) = {\log\lambda_+(\tau)\over\log a},
\end{equation}
where we use the subscript $a$ to distinguish this function from (4).

One easily sees from eq.(5) that the transfer matrix
provides information on the scale of successive splittings of
elements of a given set at every step of its construction.
It should be clear by now how the problem maps onto a one-dimensional
Ising model. The number of spin states at each site of the
corresponding Ising chain is equal to $a$. The range of interaction
depends on how far back the memory goes in determing
daughter-to-mother ratios. If we truncate
$\e_{n+1}, \e_n, ..., \e_1$ after $\e_n$ then we have
nearest-neighbor interaction. Such truncation is justified by the
fact that in many sets (in particular, that belong to the boderline
of chaos) the memory usually falls off exponentially.

Therefore, a strategy for the extraction of dynamic information from
multiplicity data is as follows. Taking the
$\tau(q)$ dependence from experiment we solve eq.(10) in the
lowest-order nontrivial case of $2 \times 2$ transfer matrix.
Writing the characteristic polynomial
\begin{equation}
\lambda^2(\tau) - \lambda(\tau) (\s_{00}^{-\tau} + \s_{11}^{-\tau})
+ (\s_{00} \s_{11})^{-\tau} - (\s_{01} \s_{10})^{-\tau} = 0,
\end{equation}
one readily sees that $\s_{01}$ and $\s_{10}$ appear only as a product,
and thus $\lambda(\tau)$ depends on three scales. Hence, together with
$a$ we have four unknowns and, therefore, need at least four
experimental points $\tau(q)$.

Usually, one of them is taken at
$q=0$, since $\tau(0) = -D_0$, where $D_0$ is the Hausdorff
dimension of a given set. Then, from eq.(10)
it follows that $\lambda(-D_0) = 1$ and
\begin{equation}
\s_{01} \s_{10} = [(1- \s_{00}^{D_0})(1- \s_{11}^{D_0})]^{1/D_0} .
\end{equation}

\section{Obtaining the $\tau (q)$ dependence}
The $\tau(q)$ dependence can be obtained in two ways. The first way follows
from the mentioned above method of constructing the transfer matrices
and consists in partition of all particles in a separate event,
$n_{tot}$, into $N$ distinct groups
$\Omega_i , i=1,2,..,N$ with the same number of particles, $n$, in
each group. Obviously, $nN = n_{tot}$. All such groups
$\Omega_i$ receive the same measure
$p_i = 1/N$ but different diameters (bins), $l_i$,
\begin{equation}
l_i = |y_{in} - y_{(i-1)n}| ,
\end{equation}
where $y_i$ is the rapidity of the $i$-th particle.
This way is applicable, of course, if the rapidities of all
particles in a given event are known with high enough precision.
Then, for a given $N$ we solve numerically $\Gamma(q, \tau) = 1$
thereby generating the $q(\tau)$ dependence.

The second way is dual to the first one and consists in partition
of the whole rapidity interval into equal bins of the diameter $l$.
Then different measures $p_i$ correspond to the equal
bins $l_i = l$.
More precisely, define $G_q$ moments [9],
\begin{equation}
G_q(l) = \sum_{i=1}^{N} p_i^q, \quad p_i = \frac {n_i} {n_{tot}} ,
\end{equation}
where the sum is over all nonempty bins, $n_i$ is the number of particles
in the $i$-th bin of the diameter $l$, q is real. For a range of
$l$ not too small (or, the number of empty bins is not too large),
from eq.$\Gamma(q, \tau) = 1$ it follows
\begin{equation}
G_q(l) \sim l^{\tau(q)} .
\end{equation}
This way is applied when not the rapidity of each particle
but only the number of particles in each bin is known.
Certainly, this way is less precise because the bin size
is obviously more than the measure corresponding to
particles within the bin.

Below we apply FJP method to finite sets of experimental points
regarding them as approximations to some infinite limit set.
Knowing the dependence $q(\tau)$ (via the first way) or
$\tau(q)$ (via the second way), we shall find the type of
the underlying dynamical system and corresponding transfer matrix elements.
To make sure that FJP method correctly reveals the underlying
dynamics we apply it first to a strange set well-known in the
theory of nonlinear dynamical systems, the so-called Feigenbaum
attractor (FA) [8].

\section{Limit cycles and FA}

FA is the limit fractal set for an infinite series
of period-doubling bifurcations of phase space trajectories
of dynamical systems with quadratic Poincare map. FA is similar to
the Cantor set and can be obtained by iteration of any point within
the unit interval $[0,1]$ using arbitrary quadratic map with
a unique maximum on $[0,1]$. As such map, one can use, for
example, a logistic map
\begin{equation}
x_{n+1} = \lambda x_n (1-x_n)
\end{equation}
at $\lambda = \lambda_{\i} \equiv 3.5699456...$. As
$3 < \lambda < \lambda_{\i}$ and $n \rightarrow \i$
the succession of elements ${x_n}$ converges to some limit
$2^m$-cycle ($m=1,2,...$), corresponding to a given $\lambda$.
Each such limit cycle can be regarded as a finite approximation
of FA, and corresponding $\s_{00}$, $\s_{11}$, and
$\s_{01} \s_{10}$ can be found.

We constructed the limit
$8-, 16-, 32-, 64$- and $2048$-cycles from eq.(17), see Tabl.1,
where the coordinates
of the limit superstable $8$- and $16$-cycles elements on the interval
[0,1] are listed.

\vspace{0.3cm}
Table 1.\\
\vspace{0.2cm}
\begin{tabular}{|c|c|c|} \hline
Cycle & 8-cycle & 16-cycle \\ \hline
      & .35170767 .81049147 & .34452966 .80545683 \\
$x_n$ & .37226755 .83066420 & .34787123 .80912288 \\
      & .50000000 .88114670 & .36998970 .83138061 \\
      & .54597520 .88866021 & .37853446 .83904467 \\
      &                     & .48167382 .87930049 \\
      &                     & .50000000 .88244553 \\
      &                     & .55084694 .89046898 \\
      &                     & .55888294 .89166684 \\ \hline
\end{tabular}

\vspace{0.3cm}
The corresponding $\lambda$ values are:
$\lambda_8 = 3.55464086..$, $\lambda_{16} =
3.56666738...$. The prefix ``super'' means that the given cycle
includes the element (in our case, $x = 1/2$), at which the first
derivative of the map vanishes. The $2048$-cycle reproduces
the characteristics of FA with high precision, see [3].

The nearest cycle elements have been joined in pairs
generating the partition of the full set into $N$ disjoint pieces
($N = 2^{m-1}$ for $2^m$-cycle) with the measure
$p_i = 1/N$ each. Putting different $\tau$ we obtained from eq.(4)
the $q(\tau)$ dependence for each cycle, see Fig.1.

\vspace{0.5cm}
\begin{picture}(131,100)
\put(5,5){\framebox(126,95)}
\put(5,43){\line(1,0){126}}
\put(41,5){\line(0,1){95}}
\put(22,39){-2}
\put(42,39){0}
\put(59,39){2}
\put(77,39){4}
\put(95,39){6}
\put(113,39){8}
\multiput(14,43)(9,0){13}{\line(0,1){2}}
\multiput(41,24)(0,19){4}{\line(1,0){2}}
\put(44,23){-1}
\put(44,61){1}
\put(44,80){2}
\put(126,38){q}
\put(44,92){$\tau$}
\put(8.69,5){\line(1,1){4.75}}
\put(13.44,9.75){\line(1,1){4.88}}
\put(18.32,14.5){\line(1,1){5.}}
\put(23.3,19.25){\line(1,1){5.1}}
\put(28.4,24.){\line(1,1){5.16}}
\put(33.56,28.75){\line(6,5){5.33}}
\put(38.89,33.5){\line(6,5){5.47}}
\put(44.37,38.25){\line(6,5){5.63}}
\put(50,43){\line(6,5){5.8}}
\put(55.8,47.75){\line(5,4){5.97}}
\put(61.77,52.5){\line(4,3){6.14}}
\put(67.9,57.25){\line(4,3){6.31}}
\put(74.21,62){\line(4,3){6.46}}
\put(80.67,66.75){\line(4,3){6.61}}
\put(87.28,71.5){\line(3,2){6.73}}
\put(94,76.25){\line(3,2){6.84}}
\put(100.84,81){\line(3,2){6.93}}
\put(107.77,85.75){\line(3,2){7}}
\put(114.77,90.5){\line(3,2){7.03}}
\put(121.8,95.25){\line(3,2){7.13}}
\put(11.36,5){\line(1,1){4.46}}
\put(15.8,9.75){\line(1,1){4.55}}
\put(20.35,14.5){\line(1,1){4.65}}
\put(25,19.25){\line(1,1){4.72}}
\put(29.7,24){\line(1,1){4.88}}
\put(34.58,28.75){\line(1,1){5}}
\put(39.57,33.5){\line(1,1){5.13}}
\put(44.7,38.25){\line(6,5){5.3}}
\put(50,43){\line(6,5){5.45}}
\put(55.45,47.75){\line(6,5){5.64}}
\put(61.09,52.5){\line(6,5){5.79}}
\put(66.88,57.25){\line(5,4){5.97}}
\put(72.85,62){\line(5,4){6.13}}
\put(79,66.75){\line(4,3){6.25}}
\put(85.25,71.5){\line(4,3){6.39}}
\put(91.64,76.25){\line(4,3){6.51}}
\put(98.15,81){\line(4,3){6.61}}
\put(104.8,85.75){\line(4,3){6.63}}
\put(111.4,90.5){\line(3,2){6.77}}
\put(118.2,95.25){\line(3,2){6.74}}
\put(26.38,9.75){\circle*{2}}
\put(32.1,19.25){\circle*{2}}
\put(38.46,28.75){\circle*{2}}
\put(45.8,38.25){\circle*{2}}
\put(54.73,47.75){\circle*{2}}
\put(66.04,57.25){\circle*{2}}
\put(79.35,66.75){\circle*{2}}
\put(93.6,76.25){\circle*{2}}
\put(108.3,85.75){\circle*{2}}
\put(123.05,95.25){\circle*{2}}
\put(115,87.5){16-cycle}
\put(100,93){32-cycle}
\end{picture}

\vspace{0.3cm}
Fig.1. $\tau(q)$ - dependence for the limit 16- and 32-cycles.
Experimental points (solid circles) correspond to the NA22 event
with 20 central particles, see Ch.5.

\vspace{0.5cm}
The generalized Renyi dimensions, $D_q$,
have been found then from eq.(2). The spectral function
$f(\alpha)$ is defined by the Legendre transform,
\begin{equation}
\alpha (q) = \frac {d \tau} {dq} , \quad f(\alpha (q)) = q \alpha (q) - \tau
(q).
\end{equation}

The maximal value of $f(\alpha)$,
$f_{max}$, is equal to the Hausdorff dimension, $D_0$. The end points
of the curves (for which $f(\alpha) = 0$),
$\alpha_{min}$ and $\alpha_{max}$,
are equal to $D_{\i}$ and $D_{-\i}$, respectively. In Table 2,
the obtained $D_{\i}$, $D_0$ and $D_{-\i}$ values are listed.

\newpage
Table 2.\\
\vspace{0.2cm}
\begin{tabular}{|c|c|c|c|c|c|} \hline
$2^m$ & 8 & 16 & 32 & 64 & $\infty$ \\ \hline
$D_{\infty}$ & .284 & .310 & .324 & .334 & .37775... \\
$D_0$ & .358 & .403 & .430 & .448 & .537... \\
$D_{- \infty}$ & .446 & .519 & .563 & .593 & .75551... \\ \hline
a & 2.0002 & 2.0000 & 1.9999 & 1.9999 & 2.0000... \\ \hline
$\s_{00}$ & .2091 & .2613 & .2852 & .3029 & .3995... \\
$\s_{11}$ & .0857 & .1062 & .1160 & .1234 & .1596... \\
$\s_{01} \s_{10}$ & .0210 & .0315 & .0404 & .0458 & .0722... \\ \hline
$-ln2/ln \s_{11}$ & .282  & .309  & .322  & .331  & .37775... \\
$-ln2/ln \s_{00}$ & .443  & .517  & .553  & .580  & .75551... \\ \hline
\end{tabular}

\vspace{0.3cm}
Remind that $D_{-\i}$ is determined by the most rarefied intervals in the set
whereas $D_{\i}$ - by the most concentrated ones.
Designate the lengths of these intervals as
$l_{-\i}$ and $l_{\i}$, respectively. As has been shown by
Feigenbaum [5], in FA they have scales
\begin{equation}
l_{-\i} \sim \alpha_F^{-n} , \quad l_{\i} \sim \alpha_F^{-2n},
\end{equation}
where $\alpha_F$ = 2.50290787... is the universal Feigenbaum
constant for quadratic maps. Then the limit dimensions can be
calculated analytically [3]:
\begin{equation}
D_{-\i}^{FA} = \frac {ln2} {ln \alpha_F} = 0.755512..,
\end{equation}
\begin{equation}
D_{\i}^{FA} = \frac {ln2} {ln \alpha_F^2} = 0.377756..,
\end{equation}
\begin{equation}
D_{-\i}^{FA} = 2D_{\i}^{AF}.
\end{equation}
Knowing the $q(\tau)$ dependence, from eqs.(10) and (12) we find
the elements of the corresponding transfer matrix and the value
of $a$, see Table 2. Just as we expected, $a$ equals $2$ with high accuracy
for all cycles, whereas the values of
$\sigma_{00}, \sigma_{11}$ and
$\sigma_{01} \sigma_{10}$ converge from below towards the limit FA values.
By definition, see eq.(5), these limit $\sigma_{ij}$ values
can be written using the Feigenbaum constant,
\begin{equation}
\sigma_{00}^{FA} = \alpha_F^{-1}, \quad \sigma_{11}^{AF} = \alpha_F^{-2}.
\end{equation}
Then the Renyi dimensions are related to
$\sigma_{00}$ and $\sigma_{11}$ as following
\begin{equation}
D_{-\i}^{FA} = - \frac {ln2} {ln \sigma_{00}}, \quad
D_{\i}^{FA} = - \frac {ln2} {ln \sigma_{11}}.
\end{equation}
These equalities are valid with high accuracy for the limit cycles
as well, compare lines 2,4 and 9,10 in Table 2.

Thus, FJP method allows us to unambiguously reveal the fractal nature
of the limit cycles:
\begin{description}
\item[-] they are self-similar if the refinement doubles successively
($a = 2$);
\item[-] two scales ($\s_{00}$ and $\s_{11}$) take part in their
construction, with these scales converging towards
$\alpha_F^{-1}$ and $\alpha_F^{-2}$, respectively;
\item[-] no element of $\s$-matrix is zero.
\end{description}

The only dynamics consistent with this is that of infinitely
period-doubling bifurcations (PDB) generated by Poincare map with
quadratic maximum. If some other fractal shows the same properties
of the transfer matrix elements then one may be sure that the
dynamics underlying that fractal is that of PDB.

One should remark that as we have seen the finite sets give the values
of $q_{exp}(\tau)$
and $D_q$ different from the limit ones, and diameter fluctuations
$l_i\rightarrow l_i+\delta l$ cause $q$-fluctuations,
$q\rightarrow q + \delta q$, where
\begin{equation}
\delta q \sim -\tau\delta l {N^{q(\tau+1)-q(\tau)}\over\log N}.
\end{equation}
Hence, to reduce the influence of experimental fluctuations
on the final result at least some $\tau$ values should be chosen around zero.

Really, the study of the limit $8,16,32,64$-
and $128$-cycles shows that for
$-1\leq\tau\leq 1$ and $a=2$ eq.(11) gives the
$q_a(\tau)$-value fitting the ``experimental'' function
$q(\tau)$ from eq.(4) up to $10^{-3}$. The coincidence of these
two functions, $q(\tau)$ and $q_a(\tau)$, for some finite set
is a reliable justification for applying FJP method.

Note that eqs.(2) and (10) for $q=1$ read $a= \lambda (0)$.
For a $2 \times 2$ matrix, eq.(12) allows only two solutions,
$\lambda_+ = 2$ if all $\s_{ij} \neq 0$; and
$\lambda_+ = (\sqrt 5 +1)/2$ if $\s_{00} = 0$ or $\s_{11} =0$.
So the obtained value $a = 2.000$ (see Table 2) is a very nontrivial
result even if one uses $\tau(q)$ values around zero.
As a matter of fact, it is very nontrivial that a solution does exist,
though a system of four nonlinear equations is not obliged to have it at all!

\section{Comparison with experiment}
For comparison we have chosen two experimental events: the well-known
$NA22$ anomalous event (AE) in $\pi^+ p$ interaction at $\sqrt s = 22 GeV$
[10] and AE observed by $JACEE$
Collaboration in $Si-AgBr$ interaction at the energy $4.1 \pm 0.7$
$TeV/nucleon$ (cosmic rays) [11]. These AE were analyzed, in
particular, by one of us (A.V.B.) in ref.[7] where for the first time
it was noted that the arrangement of particles on the rapidity axis
resembled the arrangement of the relevant limit $2^m$-cycle elements.

We obtained the $\tau(q)$ dependence for AE [10] in the first way
while for AE [11] - in the second one (via $G_q$ moments) because
in this case only hystogramme $dN/d \eta$ was measured at
pseudo-rapidity resolution $\delta \eta = 0.1$. For AE [11], not
the whole hystogramme was analyzed but only its excess over
smooth background [12,7] - 119 charged particles out of the total number
$n_{ch} = 1010 \pm 30$. The $\tau(q)$-dependence for the NA22 event
is plotted in Fig.1, whereas the relevant Renyi
dimensions and transfer matrix elements are listed in Table 3.

\vspace{0.3cm}
Table 3.\\

\vspace{0.2cm}
\begin{tabular}{|c|c|c|c|c|c|c|c|} \hline
$n_{tot}$ & 26 & 24 & 22 & 20 & 18 & 16 & 119 \\ \hline
$D_{\infty}$ & .337 & .326 & .315 & .302 & .288 & .273 & .487 \\
$D_0$ & .811 & .666 & .737 & .568 & .657 & .492 & .782 \\
$D_{- \infty}$ & 1.241 & 1.181 & 1.163 & .975 & 1.066 & .881 & 1.37 \\ \hline
a & 2.000 & 1.999 & 1.999 & 2.000 & 2.000 & 2.000 & 1.999 \\ \hline
$ \s_{00}$ & .550 & .558 & .579 & .470 & .608 & .458 & .609 \\
$ \s_{11}$ & .125 & .117 & .108 & .098 & .088 & .076 & .241 \\
$ \s_{01} \s_{10} $ & .238 & .120 & .166 & .089 & .101 & .049 & .140 \\ \hline
$-ln2/ln \s_{11}$   & .334 & .324 & .312 & .299 & .286 & .270 & .488 \\
$-ln2/ln \s_{00}$   & 1.16 & 1.19 & 1.27 & .918 & 1.39 & .888 & 1.40 \\ \hline
\end{tabular}

\vspace{0.5cm}
The last column ($n_{tot} = 119$) corresponds to AE [11] and
the $2-6$th columns correspond to AE [10]. The numbers in the first
line refer to the number of particles included into the analysis.
We successively removed two particles at each step (the last particle
from the forward and backward hemisphere, respectively) picking out
the central region of interaction where the most striking
manifestation of PDB was expected [6,7].

For both events FJP method revealed the dynamics of doublings
($a=2$) and no element of $\sigma$-matrix was zero. Also quite
satisfactory coincidence was found between theoretical and experimental
values of the quantity $\s_{11}$ (and, of course, the values of $D_{\i}$,
cf. Tables 2 and 3) that is responsible for the most concentrated
intervals of a given fractal object.

Some discrepancy in the values of
$\s_{00}$ responsible for the least dense intervals of a fractal
can be attributed to the decays of clans into secondaries
which ``rarefy'' the true arrangement of the clans on the rapidity axis.
This is confirmed also by the values of $D_{-\i}, D_0$
and $D_{\i}$ for AE [11], which practically coincide with the corresponding
averaged values for the events in ultra-relativistic heavy ion collisions [13].

It should be noted that eq.(24) remains valid for experimental
events, particularly for $D_{\i}$ and $\s_{11}$, not distorted by
the clan decays. This provides reason to suggest that transfer matrix
elements can be obtained immediately from the known
$D_{\i}, D_0$ and $D_{-\i}$ from eqs.(24) and (13), without solving
the system of equations (12).

\section{Analogy with spin-spin interaction in magnetic field}
Developing an analogy with a one-dimensional nearest-neighbour Ising model
we write the relevant action, $S$,
\begin{equation}
S = -J \sum_{i} \nu_i \nu_{i+1} - K \sum_{i} \nu_i \quad,
\end{equation}
where $\nu_i$ is a spin variable at the $i$-th site taking on the values
$0$ and $1$, $J= \beta E$, $K= \beta H$, $\beta$ is the inverse temperature,
$E$ is the spin-spin coupling constant and
$H$ is the applied magnetic field. Then the corresponding transfer
matrix looks as follows
\begin{displaymath}
\sigma = \xi \left( \begin{array}{cc}
e^{-J-K} & e^J \\
e^J & e^{-J+K} \\
\end{array} \right)
\end{displaymath}
where $\xi$ is the normalization factor. Then we get
\begin{equation}
\xi = (\s_{00} \s_{01} \s_{10} \s_{11})^{1/4} , \quad J = \frac {1} {4} ln
\frac {\s_{01} \s_{10}} {\s_{00} \s_{11}} , \quad K = \frac {1} {2}
ln \frac {\s_{11}} {\s_{00}} .
\end{equation}
In particular, in the case of FA we get
\begin{equation}
K = - \frac {1} {2}ln \alpha_F = -0.458726.., \quad J + 3K = ln \xi .
\end{equation}
In Fig.2 the pairs ($K,J$) are plotted which correspond to
the limit $2^m$-cycles, FA, NA22 event [10] and JACEE event [11].

\vspace{0.3cm}
\begin{picture}(105,60)
\put(10,5){\framebox(96,52)}
\put(10,23){\line(1,0){96}}
\put(58,5){\line(0,1){2}}
\put(8,0){-1}
\put(56,0){-0.5}
\put(104,0){0}
\put(10,41){\line(1,0){2}}
\put(3,57){0.2}
\put(3,41){0.1}
\put(3,23){0}
\put(3,5){-0.1}
\put(3,50){J}
\put(84,1){K}
\put(62,20){\circle*{2}}
\put(45,19){JACEE}
\put(63,31){\circle*{1}}
\put(63.5,29.5){\circle*{1}}
\put(63,28){\circle*{1}}
\put(63,26){\circle*{1}}
\put(62.5,28){\circle*{1.5}}
\put(56,27){FA}
\put(65,32){limit}
\put(65,29){cycles}
\put(31,51){\circle*{2}}
\put(26,46){NA22}
\end{picture}

\vspace{0.3cm}
Fig.2. Pairs ($K,J$) from eq.(26), corresponding to limit
$2^m$-cycles, Feigenbaum attractor, NA22 event [10] and JACEE event [11].
For NA22 event, 20 central particles were included in the analysis; for
JACEE event, 119 particles over a smooth background, see the text.

\vspace{0.5cm}
The points corresponding to the limit cycles and FA practically
coincide (with slight difference in $J$), very close is also the
point corresponding to JACEE event. The ``$NA22$-point'' is somewhat
away from other points. It is interesting that in terms of
$K,J$ we have very weak spin-spin interaction
and strong negative magnetic field  ($K \simeq -15 J$) for the limit
cycles, FA and JACEE event.

In conclusion of this chapter it should be emphasized that
our interpretation of the relation between multiplicity
distributions in particle collisions and a one-dimensional Ising
model is drastically different from the interpretation
accepted in literature [14-16]. Conventionally bins on the rapidity
axis are identified with sites on the Ising chain. At each site,
a spin variable $\psi$ is defined, with $\psi = 1$ if there is a particle
in the corresponding bin, and $\psi = 0$, if not.
In this case, a transfer matrix reflects interaction of such spins.
In our model, no spin variables are attributed to particles,
the former appear only at transitions between different partitions
{\em of the same} set of particles that is regarded as an
approximation of some {\em infinite} fractal set. Thus,
``spin-spin interaction'' in our approach  is in no way related
to the observed rapidity correlations of charged particles.

\section{Conclusion}
In this paper for the first time in hadron physics we have applied
FJP transfer matrix method known earlier and applied earlier
in fluid mechanics to analyze the nature of turbulence.
Using FJP method in the analysis of two anomalous events
in $hh$ collisions we managed to unambiguously reveal the
dynamics generating the given rapidity distributions of particles.
It turned out to be the dynamics of period-doubling bifurcations
characterized by Feigenbaum universality well studied during recent years.
The hypothesis on this universality was proposed long ago by one
of us (A.V.B.) so it was a pleasure to find rigorous confirmation
of this hypothesis.

Thus, we are able now to calculate analytically
many characteristics of hadron events
[6,17]: the exponent of the rise of the mean multiplicity in different
rapidity intervals, the ratios of lengths and heights of ``steps''
in the multiplicity dependence of the mean transverse momentum
at fixed energy of collisions, etc.
An analogy with Ising model in FJP method appears at successive
partitions of fractal object, so that the ``spin-spin interaction''
plays the role of topological characteristic in no way being
related to correlations of real particles in real phase space.
FJP method is similar to using simultaneously several values of
spectral function that allows one to extract dynamical information
from static description of the object under consideration.

This work was supported by Soros
International Science Foundation (A.V.B. and S.M.S.) and by the Grant of
Moscow Physics Society (S.M.S.). We thank Profs.P.V.Chliapnikov,
A.K.Likhoded, G.P.Pron'ko and Yu.G.Stroganov
for many discussions.

\end{document}